\documentclass[a4paper,UKenglish,cleveref, autoref, thm-restate]{lipics-v2021}

\pdfoutput=1 
\hideLIPIcs  

\title{Optimal Layout Synthesis for Deep Quantum Circuits on NISQ Processors with 100+ Qubits
\thanks{This is the technical report for
I. Shaik and J. van de Pol, {\em Optimal Layout Synthesis for Deep Quantum Circuits on NISQ Processors with 100+ Qubits}.
 In: Proc. IC on Theory and Applications of Satisfiability Testing, {SAT} 2024, 
(SAT'24), Pune, India, 2024.}}

\titlerunning{OLS for Deep QC on 100+ Qubit NISQ Processors} 

\author{Irfansha Shaik}{Department of Computer Science, Aarhus University, Denmark\\Kvantify Aps, DK-2300 Copenhagen S, Denmark}{irfansha@cs.au.dk}{0000-0002-7404-348X}{}
\author{Jaco van de Pol}{Department of Computer Science, Aarhus University, Denmark}{jaco@cs.au.dk}{0000-0003-4305-0625}{}

\authorrunning{Irfansha Shaik and Jaco van de Pol}

\Copyright{Irfansha Shaik and Jaco van de Pol}

  

\supplement{}
\supplementdetails[subcategory={}, cite={}, swhid={}]{Software}{https://github.com/irfansha/Q-Synth}
\supplementdetails[subcategory={Github Tag}, cite={}, swhid={}]{Software}{https://github.com/irfansha/Q-Synth/releases/tag/Q-Synth-v2.0-SAT2024}

\nolinenumbers 

\EventEditors{Supratik Chakraborty and Jie-Hong Roland Jiang}
\EventNoEds{2}
\EventLongTitle{27th International Conference on Theory and Applications of Satisfiability Testing (SAT 2024)}
\EventShortTitle{SAT 2024}
\EventAcronym{SAT}
\EventYear{2024}
\EventDate{August 21--24, 2024}
\EventLocation{Pune, India}
\EventLogo{}
\SeriesVolume{305}
\ArticleNo{22}

\usepackage{amsmath}
\usepackage{amssymb}
\usepackage{booktabs,multirow}
\usepackage{todonotes}

\usepackage{mathtools}

\usepackage{amsthm}
\usepackage{tikz}

\usepackage{subcaption}
\usepackage{amsfonts}
\usepackage[qm]{qcircuit}
\usepackage{textcomp}
\usepackage{xcolor}
\usepackage{multirow}

\usepackage{newfloat}
\usepackage{listings}

\usepackage{amsfonts}
\usepackage{algorithm}
\usepackage[noend]{algorithmic}
\usepackage[qm]{qcircuit}
\usepackage{textcomp}
\usepackage{xcolor}
\usepackage{multirow}

\usepackage{newfloat}
\usepackage{listings}

\DeclareMathOperator{\EO}{ExactlyOne}
\DeclareMathOperator{\AO}{AtmostOne}
\DeclareMathOperator{\ET}{ExactlyTwo} 
\DeclareMathOperator{\mapvar}{m}

\DeclareMathOperator{\swapvar}{s}
\DeclareMathOperator{\mappedp}{mp}

\DeclareMathOperator{\cnotvar}{c}
\DeclareMathOperator{\acvar}{ac}
\DeclareMathOperator{\dcvar}{dc}
\DeclareMathOperator{\assumvar}{asm}

\DeclareMathOperator{\swaponevar}{st}
\DeclareMathOperator{\logicalpairvar}{lp}

\DeclareMathOperator{\lqubits}{L}
\DeclareMathOperator{\pqubits}{P}
\DeclareMathOperator{\swaps}{S}

\DeclareMathOperator{\numlqubits}{nl}
\DeclareMathOperator{\numpqubits}{np}
\DeclareMathOperator{\numcnots}{nc}

\DeclareMathOperator{\cppairs}{CP} 
\DeclareMathOperator{\clpairs}{CL} 

\DeclareMathOperator{\dictcnotlq}{\mathcal{D}} 

\begin{document}

\maketitle

\begin{abstract}
  Layout synthesis is mapping a quantum circuit to a quantum processor.
  SWAP gate insertions are needed for scheduling 2-qubit gates only on connected physical qubits.
  With the ever-increasing number of qubits in NISQ processors, scalable layout synthesis is of utmost importance.
  With large optimality gaps observed in heuristic approaches, scalable exact methods are needed.
  While recent exact and near-optimal approaches scale to moderate circuits, large deep circuits are still out of scope.
  In this work, we propose a SAT encoding based on parallel plans that apply 1 SWAP and a group of CNOTs at each time step.
  Using domain-specific information, we maintain optimality in parallel plans while scaling to large and deep circuits.
  From our results, we show the scalability of our approach which significantly outperforms leading exact and near-optimal approaches (up to 100x).
  For the first time, we can optimally map several 8, 14, and 16 qubit circuits onto 54, 80, and 127 qubit platforms with up to 17 SWAPs.
  While adding optimal SWAPs, we also report near-optimal depth in our mapped circuits.
\end{abstract}

\section{Introduction}
\label{sec:introduction}

The Quantum Layout Mapping problem takes as input a quantum circuit (logical design)
and a coupling map (connectedness between physical qubits). The result is an
``equivalent'' quantum circuit mapped to the physical qubits, such that any
binary operation only happens on connected qubits. Besides an initial mapping of
logical qubits to physical qubits, this also involves the insertion of SWAP gates.
Noise is inherent to qubits in Noisy Intermediate-Scale Quantum (NISQ) processors.
Additional SWAP gates increase both the 2-qubit gate count and the circuit depth.
In the current NISQ era, minimizing error is of utmost importance for any practical quantum computing.
The error rate depends on the number of gates, the fidelity of gates, and the depth of the circuit.
The Optimal Quantum Layout Synthesis is to synthesize a mapping that optimizes one of the above metrics.

Optimal Layout Synthesis has been studied before.
A nice overview is provided in \cite{DBLP:journals/tc/TanC21}.
Several heuristic approaches exist which optimize various metrics.
The classical algorithm for heuristic mapping is SABRE (in Qiskit) \cite{DBLP:conf/asplos/LiDX19}.
\cite{DBLP:conf/glvlsi/SrivastavaLCAB23} use the MQT benchmarks 
for mapping and swapping, using a heuristic search space reduction with an $O(n\log{n})$ algorithm.
Other approaches used include A* with cost metrics \cite{DBLP:conf/rc/ZulehnerBW19}, MAXSAT \cite{DBLP:conf/micro/MolaviXDPTA22}, 
temporal planning \cite{ijcai2017p620}, and constraint programming~\cite{booth2023constraint} (minimizing circuit depth).

While heuristic approaches are fast and scalable, their suboptimal mappings may result in
high error rates~\cite{DBLP:conf/dac/WilleBZ19,DBLP:journals/tc/TanC21}.
Optimizing fidelity with exact approaches can result in circuits with the lowest error rate.
However, as shown in~\cite{DBLP:conf/iccad/TanC20}, optimizing fidelity is extremely hard
and does not scale beyond small circuits.
Circuit depth and 2-qubit gate count optimization are better alternatives for scalability.
The OLSQ tool\footnote{OLSQ tool \url{https://github.com/tbcdebug/OLSQ}} optimizes circuit depth and
is built on \cite{DBLP:conf/iccad/TanC20}.
A scalable variant OLSQ2 based on Z3 appeared in~\cite{OLSQ2_2023}.
The QMAP tool\footnote{Munich Quantum Toolkit QMAP \url{https://github.com/cda-tum/qmap}} optimizes the number of SWAP gates and
is based on \cite{DBLP:journals/tcad/ZulehnerPW19, DBLP:conf/dac/WilleBZ19}.
The same authors introduced the use of subarchitectures \cite{Peham_2023}.
Other ideas to improve quantum layout use quantum teleportation \cite{DBLP:conf/aspdac/HillmichZW21}.
In \cite{brandhofer2023optimal}, measurements are placed early so qubits can be reused.

In~\cite{ShaikvdP2023}, we proposed a tool, Q-Synth v1\footnote{Q-Synth v1 tool \url{https://github.com/irfansha/Q-Synth/releases/tag/Q-Synth-v1.0-ICCAD23}}, for SWAP gate optimization which outperformed both QMAP and OLSQ tools.
Q-Synth v1 reduces optimal quantum layout synthesis to classical planning.
For maintaining the optimality of the SWAP gates added, Q-Synth v1 adds exactly 1 CNOT or 1 SWAP gate at each time step.
In such an approach, the hardness increases with the plan length i.e., the number of CNOTs + SWAPs.
Despite the recent progress in Q-Synth v1 and OLSQ2, deep circuits that require many SWAPs are still out of reach.

\paragraph*{Contribution}
In this paper, we provide a SAT encoding based on parallel plans with domain-specific information.
In particular, at each time step, we map one SWAP gate and a group of CNOT gates.
This reduces the make-span, and using domain-specific information we maintain the optimality.
We propose two-way constraints for CNOT dependencies for better dependency propagation.
In addition, we also provide variations of our encoding with bridges and relaxed dependencies (via commutation).
In all variations, we only add provably optimal number of bridges+SWAPs.

For experimental evaluation, we consider two benchmark sets: 1) Standard benchmarks from previous papers; and 2) Deep VQE benchmarks.
For comparison, we consider leading near-optimal tool TB-OLSQ2~\cite{OLSQ2_2023} and heuristic SABRE~\cite{DBLP:conf/asplos/LiDX19}.
For mapping, we consider 4 NISQ processors Melbourne (14 qubits), Sycamore (54 qubits), Rigetti (80 qubits), and Eagle (127 qubits).
We propose three experiments: in the first two experiments we map both benchmark sets to the Sycamore, Rigetti, and Eagle platforms.
In the first experiment, we compare the number of SWAPs added by all three tools.
In the second experiment, we compare SWAP additions and circuit depth of the mapped circuits with TB-OLSQ2.
In the third experiment, we compare the effectiveness of bridges and relaxed dependencies in our tool by mapping onto the Melbourne platform.
Here we report the additional number of (optimal) SWAPs+bridges.

We demonstrate that our encoding can optimally map deep circuits onto large platforms with up to 127 qubits.
Our tool outperforms the leading near-optimal tool TB-OLSQ2 up to 100x while always adding the optimal number of SWAPs.
We show that while adding optimal SWAPs, we also report near-optimal depth in the mapped circuits.
We also confirm that heuristic approaches like SABRE add too many SWAPs.

\section{Preliminaries}
\label{sec:Preliminaries}

\subsection{Layout Synthesis for Quantum Circuits}
\label{subsec:layoutsynthesis}
A quantum circuit consists of a fixed number of (logical) qubits, and a number of quantum gates (operations) 
that are applied to some qubits in a particular order. If the output qubit of gate $g_1$ is used as an 
input qubit of gate $g_2$, we say $g_2$ depends on $g_1$. The dependencies form a DAG (directed acyclic graph)
between the gates. Gates that are (transitively) independent are called parallel, and can be applied in any order.

Any quantum circuit can be decomposed to an intermediate representation with only single-qubit gates and CNOT gates~\cite{Cross2022}.
Viewed classically, the binary CNOT gate (controlled-NOT, also known as CX) takes two qubits $(a,b)$ as input and transforms them
into $(a,a\oplus b)$, i.e., the control qubit $a$ determines whether the data qubit $b$ is negated. We will also use the SWAP
gate, which transforms a qubit pair $(a,b)$ into $(b,a)$. A SWAP gate can be expressed as a sequence of 3 CNOT gates.

The single-qubit gates will be treated as black-boxes in this paper; they
are distinguished by their name (X, Z, H, S, T, etc.) but we don't make assumptions
on their semantics (except in an extension of our method in Section~\ref{subsec:additinalfunc}).
We refer the interested reader to~\cite{Nielsen_Chuang_2010} for a detailed introduction to quantum gates and quantum circuits in general.

Most physical quantum platforms have limited connectivity, in which the CNOT operations can only be applied on
physical qubits that are neighbors in the so-called coupling graph.
Given such a circuit and a coupling graph, Layout Synthesis consists of two phases: Initial Mapping and Qubit Routing.
In Initial Mapping, the logical qubits of the given circuit are mapped to some physical qubits of the platform bijectively.
In Qubit Routing, the following constraints must be satisfied:
\begin{itemize}
  \item Every gate must be scheduled in an order that respects all dependencies;
  \item Every gate must be applied to the correct qubits (taken the mapping into account);
  \item The 2-qubit CNOT gates can only be mapped on connected physical qubits.
\end{itemize}
Additional SWAP gates may be required, to swap the values of connected physical qubits to ensure all CNOT gates can be mapped.
In this paper, we use gate count as an optimization metric.
In Layout Synthesis, the number of single-qubit gates and CNOT gates remain unchanged.
Optimal Layout Synthesis is thus minimizing the additional SWAP gates.

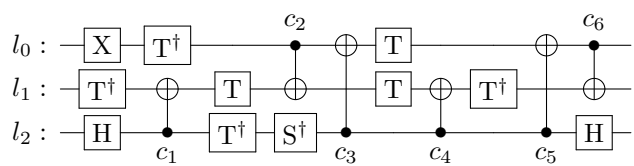
\begin{figure}[htbp]
\centering
\scalebox{1}{
  \Qcircuit @C=0.7em @R=0.2em @!R { \\
  &  & & & \dstick{c_2} & & & & & & \dstick{c_6}\\
  \lstick{{l}_{0} :} & \gate{\mathrm{X}} & \gate{\mathrm{T^\dagger}} & \qw & \ctrl{1} & \targ & \gate{\mathrm{T}} & \qw & \qw & \targ  & \ctrl{1} & \qw\\
  \lstick{{l}_{1} :} & \gate{\mathrm{T^\dagger}} & \targ & \gate{\mathrm{T}} & \targ & \qw & \gate{\mathrm{T}} & \targ & \gate{\mathrm{T^\dagger}} & \qw & \targ & \qw\\
  \lstick{{l}_{2} :} & \gate{\mathrm{H}} & \ctrl{-1} & \gate{\mathrm{\mathrm{T^\dagger}}} & \gate{\mathrm{S^\dagger}} & \ctrl{-2} & \qw & \ctrl{-1} & \qw &\ctrl{-2} & \gate{\mathrm{H}} & \qw\\
  &  & \ustick{c_1}  &  & & \ustick{c_3} & & \ustick{c_4} & & \ustick{c_5} &
  }}
  \caption{$3$-qubit Or circuit with $6$ CNOT gates.}
  \label{fig:orcircuit}
\end{figure}

\begin{figure}[thbp]
  \centering
  \begin{subfigure}{0.6\textwidth}
    \centering
    \scalebox{1}{
      \Qcircuit @C=1em @R=0.3em @!R { \\
      & &      & c_2 & & & & c_6\\
      \nghost{} &\lstick{{l}_{0} :} & \qw          & \ctrl{1}  & \targ & \qw & \targ            & \ctrl{1} & \qw\\
      \nghost{} &\lstick{{l}_{1} :} & \targ        & \targ      & \qw & \targ & \qw              & \targ & \qw\\
      \nghost{} &\lstick{{l}_{2} :} & \ctrl{-1}     & \qw        & \ctrl{-2} & \ctrl{-1} & \ctrl{-2} & \qw & \qw\\
      & & c_1 & & c_3 & c_4 & c_5 & & 
       }}
       \caption{Or-circuit with only CNOT gates.}
       \label{fig:orcnotcircuit}
  \end{subfigure}%
  \begin{subfigure}{0.4\textwidth}
    \centering
    \begin{tikzpicture}
      \node[shape=circle,draw=black] (A) at (0,0) {$p_0$};
      \node[shape=circle,draw=black] (B) at (1,1) {$p_1$};
      \node[shape=circle,draw=black] (C) at (2,0) {$p_2$};
      \path [-] (A) edge (B);
      \path [-](B) edge (C);
    \end{tikzpicture}
    \caption{Coupling graph.}
    \label{fig:couplinggraph}
  \end{subfigure}
  \caption{Reduced Or circuit and a $3$-qubit linear platform.}
  \label{fig:orcnotcircuitandcp}
\end{figure}
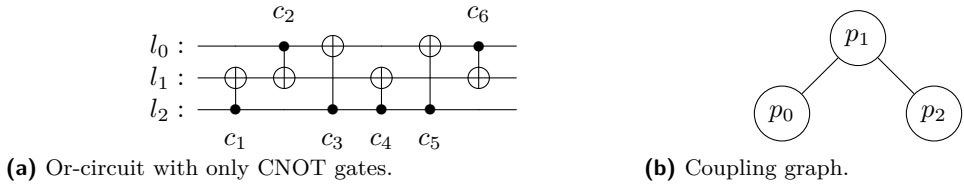
For example, Figure~\ref{fig:orcircuit} shows an Or-circuit with $3$ logical qubits (horizontal lines $\{l_0, l_1, l_2\}$).
The circuit has $11$ single-qubit gates (boxes with names) and $6$ CNOT gates (the dot indicates the control qubit, while the $\oplus$ indicates the data qubit).
Let us suppose we want to map this circuit onto the linear 3-qubit platform as in Figure~\ref{fig:couplinggraph}.
Regardless of physical qubit connections, single-qubit gates can always be scheduled.
Only the 2-qubit CNOT gates are relevant for our optimal synthesis problem.
Thus, we first remove the single-qubit gates and only consider CNOT gates for the mapping. After finding the optimal mapping, the single-qubit gates will
be reinserted.
Figure~\ref{fig:orcnotcircuit} shows the CNOT gates in the Or-circuit.

In any valid mapping, the dependencies must be respected, for example, gates $c_1$ and $c_3$ can only be mapped before and after gate $c_2$, respectively.
In this example, the dependency graph is a total order, but note that with 4 qubits, parallel CNOT gates are possible,
which can be scheduled in any order.
One can observe that the connections of the CNOT gates $c_1,c_2$ and $c_3$ form a triangle $(l_0,l_1),(l_1,l_2),(l_2,l_0)$.
Since the coupling graph does not have a triangle, one cannot map our example circuit to the linear platform.
At least two SWAP gates are needed for any valid mapping.

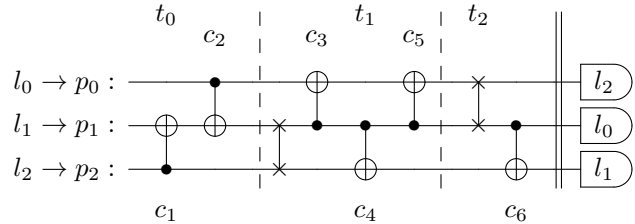
\begin{figure}[htbp]
  \centering
  \scalebox{1}{
    \Qcircuit @C=1em @R=0.3em @!R { \\
    & &  \ustick{t_0}  \ar@{--}[]+<3.5em,1em>;[d]+<3.5em,-4em>  & c_2 & &  & c_3 & \ustick{t_1} & c_5 \ar@{--}[]+<1em,1em>;[d]+<1em,-4em> & & \ustick{t_2} \ar@{=}[]+<3em,1em>;[d]+<3em,-4em> &\\
    \nghost{{l}_{2} :  } & \lstick{{l}_{0} \rightarrow {p}_{0} : } & \qw          & \ctrl{1} & \qw & \qw            & \targ    & \qw      & \targ    & \qw & \qswap \qwx[1] & \qw & \qw & \measureD{l_2}\\
    \nghost{{l}_{1} :  } & \lstick{{l}_{1} \rightarrow {p}_{1} : } & \targ        & \targ     & \qw & \qswap \qwx[1]& \ctrl{-1} & \ctrl{1} & \ctrl{-1} & \qw & \qswap & \ctrl{1}  & \qw & \measureD{l_0}\\
    \nghost{{l}_{0} :  } & \lstick{{l}_{2} \rightarrow {p}_{2} : } & \ctrl{-1}     & \qw       & \qw & \qswap       & \qw      & \targ    &  \qw     & \qw & \qw  & \targ& \qw& \measureD{l_1}\\
    & & c_1    &  & & &  & c_4 &  & & & c_6 &
     }}
     \caption{Mapped Or-circuit with 2 additional SWAPs (optimal).}
     \label{fig:cnotcircuitmapped}
\end{figure}

Figure~\ref{fig:cnotcircuitmapped} shows such a mapped Or-circuit where $l_0, l_1, l_2$ are mapped to $p_0, p_1,p_2$ respectively.
Intuitively, the SWAP gates slice the circuit such that the sub-circuits do not have any triangles by CNOT connections.

\begin{figure}[htbp]
  \centering
  \scalebox{1}{
    \Qcircuit @C=0.7em @R=0.2em @!R { \\
    \nghost{} &  &  & & & \dstick{c_2} & & & & \dstick{c_3} & & \dstick{c_5} & & \ar@{=}[]+<3em,0.5em>;[d]+<3em,-4.5em> &\\
         \nghost{{l}_{2} :  } & \lstick{{l}_{0} \rightarrow {p}_{0} :  } & \gate{\mathrm{X}} & \gate{\mathrm{T^\dagger}} & \qw & \ctrl{1} & \qw & \qw & \qw & \targ & \gate{\mathrm{T}} & \targ & \qswap & \qw & \gate{\mathrm{H}} & \qw & \measureD{l_2}\\
         \nghost{{l}_{1} :  } & \lstick{{l}_{1} \rightarrow {p}_{1} :  } & \gate{\mathrm{T^\dagger}} & \targ & \gate{\mathrm{T}} & \targ & \qswap & \gate{\mathrm{T^\dagger}} & \gate{\mathrm{S^\dagger}} & \ctrl{-1} & \ctrl{1} & \ctrl{-1} & \qswap \qwx[-1] & \qw &\ctrl{1} & \qw & \measureD{l_0}\\
         \nghost{{l}_{0} :  } & \lstick{{l}_{2} \rightarrow {p}_{2} :  } & \gate{\mathrm{H}} & \ctrl{-1} & \qw & \qw & \qswap \qwx[-1] & \gate{\mathrm{T}} & \qw & \qw & \targ & \gate{\mathrm{T^\dagger}} & \qw & \qw & \targ & \qw & \measureD{l_1}\\
         \nghost{} &  &  & \ustick{c_1}  & & & & & & & \ustick{c_4} & & & & \ustick{c_6}
     }}
      \caption{Final Mapped Or-circuit after inserting back single-qubit gates.}
      \label{fig:orcircuitmapped}
\end{figure}
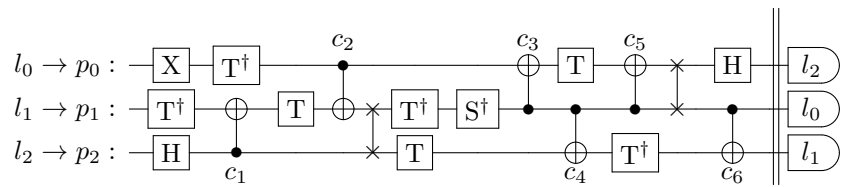

Finally, single-qubit gates can be inserted back respecting original DAG dependencies.
Figure~\ref{fig:orcircuitmapped} shows the final mapped circuit with optimal SWAP gates.
Note that the number of physical qubits can be more than logical qubits.
In such cases, one can use so-called \emph{ancillary} qubits to avoid unnecessary swaps.
Similar to Q-Synth v1, we allow ancillary swapping i.e., a mapped physical qubit can be swapped with an empty physical qubit.

\subsection{Optimal Layout Synthesis as Planning}
\label{subsec:optlsplanning}

In Q-Synth v1, we encoded optimal layout synthesis as a planning problem in the Planning Domain Definition (PDDL) Specification.
As discussed above, the reduced circuit with only CNOT gates is mapped using additional SWAPs.
Later single-qubit gates are inserted back to reconstruct the final mapped circuit.
In such a planning problem, either exactly one CNOT gate or one SWAP gate is scheduled at each time step.
A plan with the optimal number of actions corresponds to the optimal number of SWAP additions.

\paragraph*{Planning as SAT}
Given a boolean formula, a Satisfiability (SAT) problem is finding an assignment to the boolean variables that makes it a true formula.
A planning problem can be encoded as a bounded reachability problem.
Sequential encoding~\cite{Kautz1992planning} is a standard SAT encoding where each time step encodes a single action.
Using a sequential encoding, one can obtain optimal plans by incrementing the plan length by 1.
For instance, one could use Q-Synth v1 with Madagascar (a SAT-based planner) to find an optimal mapping.
Since the optimal plan length for the example, Or-circuit is 8 (6 CNOTs + 2 SWAPs), the SAT instance has a make-span of 8.
As shown in~\cite{ShaikvdP2023}, sequential encoding scales well for moderate circuits however deep circuits are still out of reach.
It is consensus that a long optimal plan length can severely impact the performance of SAT-based planners.

\paragraph*{Parallel Plans}
In literature, alternative parallel plan encodings like $\forall$-step~\cite{DBLP:conf/kr/KautzMS96} and
$\exists$-step~\cite{RINTANEN20061031} were proposed for scaling heuristic SAT-based planning.
The key idea in a parallel plan is to group two or more actions whose preconditions and effects do not conflict.
While encodings like $\exists$-step scale well, the optimality is not guaranteed.
For a scalable optimal layout synthesis, one needs a way to group CNOTs while still maintaining the optimality.

\subsection{Parallel Plans in Optimal Layout Synthesis}
\label{subsec:olsparallelplans}

Madagascar implements both $\forall$-step and $\exists$-step parallel plans.
Directly using Q-Synth v1 with parallel plan encodings in Madagascar does not preserve optimality.
In particular, there are three main challenges:
\begin{itemize}
  \item More than one SWAP gate can be applied at each parallel step;
  \item Planner needs to find a partial order in each parallel step satisfying dependencies;
  \item Relaxing CNOT dependencies within a parallel step is not trivial in a PDDL specification.
\end{itemize}

In this paper, we directly encode Layout Synthesis as a SAT problem to circumvent the encoding challenges in a PDDL specification.
In our encoding, we allow exactly one SWAP gate at each parallel step.
Thus, the number of parallel steps corresponds to the optimal SWAP additions.
Further, we use domain-specific information from Layout Synthesis to relax CNOT dependencies within a parallel step.
In particular, we make two observations:
\begin{itemize}
  \item The qubit mappings do not change between two consecutive SWAP gates;
  \item Given a set of CNOTs, one can always reconstruct a partial order with DAG dependencies.
\end{itemize}
We take advantage of the partial order reconstruction and drop dependency constraints within a parallel step.
The SAT solver can now choose exactly one SWAP and a group of CNOTs in each parallel step.
CNOT gates in different time steps still need to respect the original DAG dependencies.
The full SAT encoding is later discussed in detail in Section \ref{sec:twowaysat}.

For our example, a parallel plan with a make-span of 3 is sufficient (see Figure~\ref{fig:cnotcircuitmapped}) instead of 8.
At time step 0, along with the initial mapping, a group of CNOT gates are also mapped.
From time step 1, exactly one SWAP gate and a group of CNOT gates are mapped.
Note that the satisfying assignment returned by a SAT solver only specifies that:
\begin{itemize}
  \item Logical qubits $l_0,l_1, l_2$ are mapped to $p_0,p_1,p_2$, respectively;
  \item SWAP gates on $p_1,p_2$ and $p_0, p_1$ are applied at time steps $t_1,t_2$ respectively;
  \item $c_1, c_2$ gates are applied at $t_0$;
  \item $c_3, c_4, c_5$ gates are applied at $t_1$;
  \item $c_6$ gate is applied at $t_2$;
\end{itemize}
In mapped circuit reconstruction, we use the DAG dependencies to order the group of CNOTs in each time step.
In the literature, an SMT based encoding is applied in TB-OLSQ(2)~\cite{DBLP:conf/iccad/TanC20,OLSQ2_2023} which also groups CNOT gates between consecutive SWAP gates.
They optimize make-span of their defined problem. However, longer make-span may result in better SWAP count and circuit depth.
In our experiments, we indeed observe suboptimal solutions by TB-OLSQ2 in both metrics.

\subsection{Incremental SAT Solving}
\label{subsec:incrementalsat}

Conflict Driven Clause Learning (CDCL)~\cite{CDCL2021} is a key part of state-of-the-art SAT solving.
When solving similar instances, one can reuse the learned clauses.
Incremental SAT solving allows solving a SAT instance given an assumption of a partial assignment.
Essentially by using different assumptions, multiple instances can be solved while reusing the learned clauses.
In problems like planning, one needs to refute up to k-1 plan length for optimal plans.
By adding assumptions encoding that the goal is reached in the current iteration, 
one can solve a planning instance incrementally.

\section{Two-Way Parallel SAT encoding}
\label{sec:twowaysat}
In this section, we implement the ideas discussed above.
We provide an incremental SAT encoding that applies the idea of parallel plans in Layout Synthesis.
Table \ref{table:vars} describes the main variables used in the encoding.
Algorithm~\ref{alg1} describes the structure of our encoding.
In every time step, a group of CNOTs are applied.
From time step 1, each incremental step adds one extra SWAP.
We generate a set of variables for CNOT and SWAP constraints at each time step.

\begin{table}[htbp]
  \centering
  \caption{Encoding variables and descriptions}
  \begin{tabular}{ll}
  \hline
  Variable & Description \\
  \hline
   $\mapvar^{t}_{l,p}$  & mapping var for logical $l$ and physical $p$ qubits at time step $t$\\
   $\mappedp_{p}^{t}$ & if physical qubit $p$ is mapped to some logical qubit at time step $t$\\
   $\swapvar_{p,p'}^{t}$ & SWAP variable for physical qubits ${p,p'}$ at time step $t$\\
   $\swaponevar_{p}^{t}$ & SWAP-touched variable for physical qubit $p$ at time step $t$\\
   $\cnotvar_{i}^{t}/\acvar_{i}^{t}/\dcvar_{i}^{t}$ & current/advanced/delayed CNOT var for $i$th CNOT at time step $t$\\
   $\logicalpairvar_{l,l'}^{t}$ & logic qubit pair variables for logical qubits $l,l'$ at time step $t$\\
  \hline
  \end{tabular}
  \label{table:vars}
  \end{table}

In addition to specifying which CNOT gates are chosen in each time step, we also need to specify the CNOT dependencies.
We use the DAG generated from the original circuit for computing the CNOT dependencies.
We adapt the CNOT dependency constraints by specifying that for every CNOT gate in time step $t$ its predecessors (successors)
can be applied at time step $t'$, where $t'\leq t$ ($t'\geq t$).
We use two extra CNOT blocks, advanced and delayed CNOTs, which specify if a CNOT gate is mapped in an earlier or later time step.
We call this Two-way SAT encoding to emphasize the bidirectional propagation of CNOT dependencies.
In the following paragraphs, we describe the main parts of the Algorithm and provide the constraints.

\begin{algorithm}[htbp]
  \caption{Incremental SAT Solving, starting with t=0}
  \label{alg1}
  \begin{algorithmic}[1]
  \FORALL{$l \in [1 \dots \numlqubits]$}  \STATE $\EO(\mapvar_{l,1}^{t},\dots,\mapvar_{l,\numpqubits}^{t})$ \ENDFOR
  \FORALL{$p \in [1 \dots \numpqubits]$}
  \STATE $\AO(\mapvar_{1,p}^{t},\dots,\mapvar_{\numlqubits,p}^{t})$
  \ENDFOR
  \STATE \texttt{MappedPQubits}
  \IF{t != 0}
  \STATE \texttt{SwapConstraints} and \texttt{Ancillaries}
  \ENDIF
  \STATE \texttt{CNOTConnections} and \texttt{CNOTDependencies}
  \STATE \texttt{Assumptions}
  \STATE Solve instance with assumption $\assumvar^{t}$
  \IF {Instance not satisfied}
  \STATE repeat from step 1 with t = t + 1
  \ENDIF
  \end{algorithmic}
  \end{algorithm}

\paragraph*{Initial Mapping}
Let $\lqubits$ ($\pqubits$) be a set of logical (physical) qubits in the circuit.
Let $\numlqubits$ ($\numpqubits$) be the number of logical (physical) qubits.
In time step $0$, we add requirements on the Initial Mapping for logical and physical qubits.
Lines 1 to 4 in the Pseudo code add constraints for mapping every logical qubit to a unique physical qubit.
We use one-hot encoding for specifying the mapping. We apply ExactlyOne (AtmostOne) constraints for logical (physical) qubit mapping variables.
Adding these constraints only at the 0th time step is sufficient for correctness.
However, adding these constraints at every time step significantly improved the performance of SAT solvers.

\paragraph*{\texttt{SwapConstraints}}
From time step 1, we use the same mapping variables for handling SWAPs.
Adding a SWAP gate changes the mapping between logical and physical qubits.
Let $\cppairs$ be the set of all connected physical qubit pairs.
The following constraints must ensure that a SWAP gate is only applied on a connected physical qubit pair.
The logical qubits mapped on the physical qubit pair must be swapped in the next time step.
The qubit mappings for the untouched physical qubits must be propagated.
We define two sets of variables to satisfy such constraints.
For choosing a SWAP, we define one SWAP variable $\swapvar_{p,p'}$ for each connected physical qubit pair $(p,p')$.
For propagation, we define SWAP-touch variables $\swaponevar_{p}$ to specify if the physical qubit $p$ is touched by the SWAP.
We specify that 1) Each SWAP variable forces the SWAP-touched physical variables to True;
2) Exactly one of the SWAP variables is set to True;
3) Every SWAP forces exactly two SWAP-touched variables to True.
Let $\swaps$ be the set of all SWAP variables.
The corresponding boolean constraints are:
\begin{align*}
&\bigwedge_{(p,p') \in \cppairs} (\swapvar_{p,p'}^{t} \rightarrow \swaponevar_{p}^{t} \land \swaponevar_{p'}^{t}) \land \EO(\swaps^{t}) \land \ET(\swaponevar_1^{t}, \dots, \swaponevar_{\numpqubits}^{t})
\end{align*}

Based on the chosen SWAP variables, we update the mapping variables.
For each SWAP variable $\swapvar_{p,p'}$, we swap the $p$ and $p'$ mapped variables from the previous to the current step.
\begin{equation*}
\bigwedge_{(p,p') \in \cppairs} \bigwedge_{l \in \lqubits} \swapvar_{p,p'}^{t} \rightarrow  ((\mapvar_{l,p}^{t-1} \leftrightarrow \mapvar_{l,p'}^{t}) \land ( \mapvar_{l,p'}^{t-1} \leftrightarrow \mapvar_{l,p}^{t}))
\end{equation*}
If a SWAP-touch variable is False, we propagate the corresponding mapping variables.
\begin{equation*}
\bigwedge_{p \in \pqubits} \bigwedge_{l \in \lqubits} \neg\swaponevar_p^{t} \rightarrow (\mapvar_{l,p}^{t-1} \leftrightarrow \mapvar_{l,p}^{t})
\end{equation*}

\paragraph*{\texttt{MappedPQubits} and \texttt{Ancillaries}}
Using additional qubits, traditionally called \emph{ancillaries}, can reduce the total number of SWAPs needed. An \emph{ancillary SWAP} exchanges a mapped qubit with an unmapped qubit.
To specify this, we need to keep track of mapped qubits (including at time step $0$).
We specify that a physical qubit $p$ is mapped to some logical qubit if and only if its mapped variable $\mappedp_{p}$ is True.
\begin{equation*}
\bigwedge_{p \in \pqubits}\mappedp_{p}^{t} \leftrightarrow (\bigvee_{l\in\lqubits} \mapvar_{l,p}^{t}) 
\end{equation*}
We restrict that at least one of the swapped physical qubits is a mapped qubit.
With similar constraints, we also provide an option for only non-ancillary SWAPs.
\begin{equation*}
\bigwedge_{(p,p') \in \cppairs} \swapvar_{p,p'}^{t} \rightarrow (\mappedp_{p}^{t} \lor \mappedp_{p'}^{t}) 
\end{equation*}

\paragraph*{\texttt{CNOTConnections}}
Let $\clpairs$ be the set of all connected logical qubit pairs dervied from the CNOT connections in the input circuit.
We require CNOT gates to be applied only on connected physical qubits.
Since CNOT gates must be applied to specific logical qubits, we require that the corresponding logical qubits be mapped to connected physical qubits.
First, we specify that logical qubit pair variables are true if and only if the physical qubits they are mapped are connected.
\begin{align*}
  \bigwedge_{(l,l') \in \clpairs} \big(&\bigwedge_{(p,p') \in \cppairs} ((\mapvar_{l,p}^t \land \mapvar_{l',p'}^t) \lor (\mapvar_{l,p'}^t \land \mapvar_{l',p}^t)) \rightarrow  \logicalpairvar_{l,l'}^{t} \land \\
  &\bigwedge_{(p,p') \in \overline{\cppairs}} ((\mapvar_{l,p}^t \land \mapvar_{l',p'}^t) \lor (\mapvar_{l,p'}^t \land \mapvar_{l',p}^t)) \rightarrow  \neg\logicalpairvar_{l,l'}^{t}\big)
\end{align*}
Using logical qubit pair variables, we specify that if a CNOT is mapped then its corresponding logical qubits are connected.
We define $\dictcnotlq$ as a dictionary of CNOT indices to logical qubit pairs.
Let $\numcnots$ be the number of CNOT gates. The corresponding boolean constraint is:
\begin{equation*}
\bigwedge_{\numcnots}^{i=1} \cnotvar_{i}^{t} \rightarrow  \logicalpairvar_{\dictcnotlq[i]}^{t}
\end{equation*}

\paragraph*{\texttt{CNOTDependencies}}
For a correct mapping, we need to respect the DAG dependencies of the CNOT gates in a circuit.
As discussed earlier, we use advanced and delayed CNOT blocks in each time step to propagate local information globally.
Every CNOT is mapped, advanced, or delayed in all time steps, depending on the
status of its predecessors ($pre$) and successors ($suc$) in the dependency DAG.
If at time step $t$ a CNOT is:
\begin{itemize}
\item Mapped: Its predecessors (successors) are either advanced (delayed) or mapped in the same time step.
\item Advanced: 1) It is applied or advanced in $t-1$; 2) Its predecessors are also advanced in $t$.
\item Delayed: 1) It is either applied or delayed in $t+1$; 2) Its successors are also delayed in $t$; 3) In $t$, either its logical qubits are not connected or one of its predecessors is delayed.
\end{itemize}
The corresponding boolean constraints are:
\begin{align*}
& \bigwedge_{\numcnots}^{i=1} \big(\EO(\cnotvar_{i}^t,\acvar_{i}^t,\dcvar_{i}^t) \land\\
& \quad \bigwedge_{j \in pre(i)} \cnotvar_{i}^t \rightarrow (\acvar_{j}^t \lor \cnotvar_{j}^t) \land \bigwedge_{j \in suc(i)} \cnotvar_{i}^t \rightarrow (\cnotvar_{j}^t \lor \dcvar_{j}^t) \land \\
& \quad \acvar_{i}^t \rightarrow (\cnotvar_{i}^{t-1} \lor \acvar_{i}^{t-1}) \land \bigwedge_{j \in pre(i)} \acvar_{i}^t \rightarrow \acvar_{j}^t \land\\
& \quad \dcvar_{i}^{t-1} \rightarrow (\cnotvar_{i}^{t} \lor \dcvar_{i}^{t}) \land \bigwedge_{j \in suc(i)} \dcvar_{i}^t \rightarrow \dcvar_{j}^t \land \quad \dcvar_{i}^{t} \rightarrow ( \neg \logicalpairvar_{\dictcnotlq[i]}^{t} \lor \bigvee_{j \in pre(i)} \dcvar_{i}^{t})\big)
\end{align*}

\paragraph*{Assumptions for Incremental solving}
We specify that CNOT gates cannot be advanced at time step $0$ i.e., $\bigwedge_{\numcnots}^{i=1} \neg \acvar_{i}^0$.
For using incremental solving in SAT, we use an assumption variable $\assumvar^{t}$.
At every time step, if the assumption variable is true then CNOT gates cannot be delayed i.e., $\assumvar^{t} \leftrightarrow \bigwedge_{\numcnots}^{i=1} \neg\dcvar_{i}^t$.
For each time step, we call the SAT solver with the assumption variable $\assumvar^{t}$ as True.

\paragraph*{Encoding Size}
Let $l$ be the number of logical qubits, $p$ be the number of physical qubits, $p_e$ be the number of edges in the physical coupling graph,
$l_e$ be the number of edges in the logical graph (from CNOT gates), $c$ be the number of CNOTs, and finally, let $k$ be the make-span.
The encoding requires  $O(k(lp + p_e + l_e + c))$ variables.
Usually, the physical coupling graphs are planar, so the variables required is $O(k(lp + l_e + c))$.
Note that, we use the sequential counter encoding for exactly-one constraints, so it can add extra $O(p)$ auxiliary variables.
In total, the number of variables is $O(k(lp + l_e + c))$.
The encoding requires $O(k(lp + lp_{e} + l_{e}p^2 + c))$ clauses.
Again, for a planar physical coupling graph this is bounded by $O(k(lp + l_{e}p^2 + c))$ clauses.

\subsection{Additional Functionality}
\label{subsec:additinalfunc}

So far, we have not used the semantics of the unary or binary gates (except the SWAP gates).
The previous encoding could also be used to map circuits with for instance binary CZ gates instead of CNOT gates.
If we take the semantics of the gates into account, there are more opportunities to optimize
the circuits. While a complete re-synthesis of the circuit is beyond the scope of this paper,
we want to illustrate some known techniques that can further reduce the number of SWAP gates.
We emphasize that we now change the optimization problem (by allowing more solutions), and
that the extensions are specific for CNOT gates.
Our main purpose is to show that our proposed encoding can be easily extended to 
incorporate these techniques, known as ``bridges'' and ``relaxed dependencies''. 
We also note that the encoding can be easily restricted to disallow the use of ``ancillary qubits''.

\paragraph*{Bridges}

\begin{figure}[htbp]
  \centering
  \begin{subfigure}[b]{0.4\textwidth}
    \centering
    \scalebox{0.9}{
      \Qcircuit @C=0.7em @R=0.2em @!R { \\
       &  & & & c_3 & & c_5 & & \\
      \nghost{} &\lstick{{l}_{0} \rightarrow {p}_{0} :} & \qw      & \ctrl{2} & \targ    & \qw      & \targ    & \ctrl{2}& \qw\\
      \nghost{} &\lstick{{l}_{2} \rightarrow {p}_{1} :} & \ctrl{1} & \qw      & \ctrl{-1}& \ctrl{1} & \ctrl{-1}& \qw     & \qw\\
      \nghost{} &\lstick{{l}_{1} \rightarrow {p}_{2} :} & \targ    & \targ    & \qw      & \targ    & \qw      & \targ   & \qw\\
       & & c_1 & c_2 & & c_4 & & c_6
       }}
       \caption{CNOT gates after initial mapping.}
       \label{fig:orcnotcircuitpermuted1}
  \end{subfigure}%
  \begin{subfigure}[b]{0.6\textwidth}
    \centering
    \scalebox{0.9}{
      \Qcircuit @C=0.6em @R=0.3em @!R { \\
       & \ar@{--}[]+<2.1em,1em>;[d]+<2.1em,-4em> & \ustick{t_0} &   & \ar@{--}[]+<3em,1em>;[d]+<3em,-4em>& \ustick{t_1} & c_3 & b_1 & & b_3 & \ustick{t_2} \ar@{=}[]+<3.7em,0.5em>;[d]+<3.7em,-4.5em> & c_5\\
      \nghost{} &\lstick{{l}_{0} \rightarrow {p}_{0} :}& \qw & \qw        & \qswap \qwx[1] & \qw        & \ctrl{1} & \ctrl{1} & \qw      & \ctrl{1} & \qw      & \ctrl{1} & \qw & \qw  & \measureD{l_2}\\
      \nghost{} &\lstick{{l}_{2} \rightarrow {p}_{1} :}& \ctrl{1} & \qw   & \qswap         & \ctrl{1}   & \targ    & \targ    & \ctrl{1} & \targ    & \ctrl{1} & \targ  & \ctrl{1} & \qw & \measureD{l_0}\\
      \nghost{} &\lstick{{l}_{1} \rightarrow {p}_{2} :}& \targ  & \qw    & \qw            & \targ      & \qw      & \qw      & \targ    & \qw      & \targ    & \qw & \targ & \qw & \measureD{l_1}\\
       & & c_1 & & & c_2 & & & b_2 & & b_4 & & c_6
       }}
       \caption{CNOT gates after adding a SWAP and a bridge gate.}
       \label{fig:orcnotcircuitpermuted1swapbridge}
    \end{subfigure}
  \caption{Mapping reduced circuit using a SWAP and a bridge gate.}
  \label{fig:bridgecnotcircuit}
\end{figure}
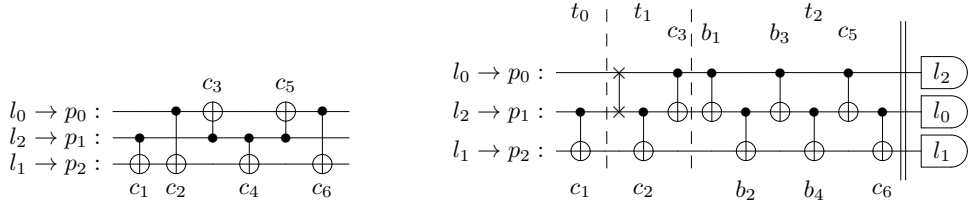

Using a bridge, one can apply a CNOT on disconnected physical qubits.
For instance, the bridge $b_1$-$b_4$ in Fig.~\ref{fig:bridgecnotcircuit}
together implements a CNOT between $p_0$ and $p_2$ (which implements $c_4$ after the preceding SWAP). 
We limit ourselves to bridges of distance 2. Observe that the bridge introduces 3 extra
CNOT gates, so the cost is the same as a SWAP gate. However, the result
is different, since a bridge does not swap the qubits. This might be an advantage,
depending on the rest of the circuit.
In~\cite{DBLP:journals/integration/ItokoRIM20}, it was shown that using bridges can reduce the overall CNOT count.

We adapted our optimal Two-Way SAT encoding, by allowing to add either a single bridge
or a single SWAP gate at each time step. If a bridge was added, the corresponding CNOT
gate is regarded as scheduled. So both options cost exactly 3 CNOT gates.
The SAT solver will find a solution with the minimal sum of bridge or SWAP gates. 
Our experiments will show that we indeed find better solutions with bridges.

\paragraph*{Relaxed Dependencies}

\begin{figure}[htbp]
  \centering
  \begin{subfigure}[b]{0.4\textwidth}
    \centering
    \scalebox{0.9}{
      \Qcircuit @C=0.7em @R=0.2em @!R { \\
       &  & c_1 & & c_3 & c_4 & c_5 & & \\
      \nghost{} &\lstick{{l}_{2} \rightarrow {p}_{0} :} & \ctrl{1}  & \qw       & \ctrl{2} & \ctrl{1} & \ctrl{2} & \qw     & \qw\\
      \nghost{} &\lstick{{l}_{1} \rightarrow {p}_{1} :} & \targ & \targ     & \qw      & \targ    & \qw      & \targ   & \qw\\
      \nghost{} &\lstick{{l}_{0} \rightarrow {p}_{2} :} &\qw        & \ctrl{-1} & \targ    & \qw      & \targ    & \ctrl{-1}& \qw\\
       & & & c_2 & & & & c_6
       }}
       \caption{CNOT gates after initial mapping.}
       \label{fig:orcnotcircuitpermuted2}
  \end{subfigure}%
  \begin{subfigure}[b]{0.6\textwidth}
    \centering
    \scalebox{0.9}{
      \Qcircuit @C=0.8em @R=0.3em @!R { \\
       & & \ustick{t_0} & \ar@{--}[]+<2.3em,1em>;[d]+<2.3em,-4em>  & c_4 & & c_3 & c_5 & \ustick{t_1} \ar@{=}[]+<1em,1em>;[d]+<1em,-4.5em> &\\
      \nghost{} &\lstick{{l}_{2} \rightarrow {p}_{0} :}& \ctrl{1}& \qw      & \ctrl{1} & \qw           & \ctrl{1}      & \ctrl{1} & \qw & \qw & \measureD{l_2}\\
      \nghost{} &\lstick{{l}_{1} \rightarrow {p}_{1} :}& \targ   &  \targ   & \targ    & \qswap \qwx[1]& \targ & \targ   & \ctrl{1} & \qw & \measureD{l_0}\\
      \nghost{} &\lstick{{l}_{0} \rightarrow {p}_{2} :}& \qw     & \ctrl{-1}& \qw      & \qswap        & \qw      & \qw & \targ & \qw & \measureD{l_1}\\
       & & c_1 & c_2 & & & & & c_6
       }}
       \caption{CNOT gates after adding a SWAP with commutation.}
       \label{fig:orcnotcircuitpermuted2relaxedswap}
    \end{subfigure}
  \caption{Mapping reduced circuit using a SWAP and relaxed dependencies.}
  \label{fig:relaxedcnotcircuit}
\end{figure}
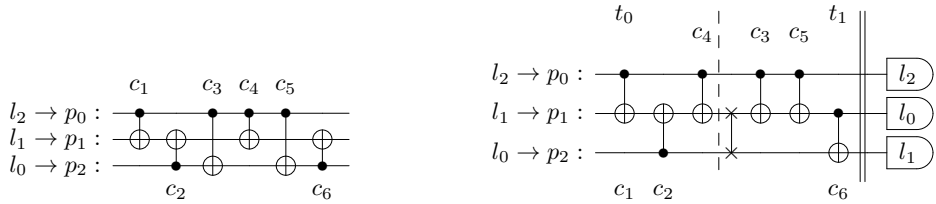

The authors of \cite{DBLP:conf/aspdac/ItokoRIMC19} consider gate commutation rules for quantum layout mapping.
Commutation and cancellation rules on $R_Z$ and CNOT gates are also used in \cite{Nam_2018}, to reduce the number of
H-gates. For instance, two subsequent CNOT gates on the same control qubit, or on the same data qubit, can be commuted without changing the semantics.
Also, single Z-like gates (like the Z-, S-, T- and $R_Z$-gates) commute with the control bit of a CNOT,
while X-like gates (like the X- and $R_X$-gates) commute with its data bit. 

This added freedom can be exploited: by permuting the CNOT gates,
they can be grouped in a convenient manner, so less SWAP gates are needed.
For instance, in Fig.~\ref{fig:orcnotcircuitpermuted2}, $c_3$ and $c_4$ can be commuted, 
since they share their control bit. As a result, we can now find a solution that requires
only 1 SWAP gate (Fig.~\ref{fig:orcnotcircuitpermuted2relaxedswap}), while still
respecting the linear coupling graph.

The gate commutation rules can be incorporated in our optimal Two-way SAT encoding
by computing a ``relaxed'' dependency graph. In the example above, we consider
$c_2$ as a dependency for $c_3$ and $c_4$, but $c_3$ and $c_4$ are considered
independent. We stress that the relaxed dependencies must also take the unary
gates into account (for instance $c_2$ and $c_4$ cannot be commuted, since there
is a T-gate in between, cf. Fig.~\ref{fig:orcircuit}).

In our tool, we compute the relaxed dependency graph before removing the unary gates.
We then generate the encoding as presented before, replacing the dependency graph 
with the relaxed dependency graph. This guarantees an optimal Layout Mapping
(minimal number of SWAPS) given the specified commutation rules.
Our experiments show that relaxed dependencies can indeed provide better solutions.

\paragraph*{Non-Ancillary Mapping}
The actual cost of ancillary qubits in practical quantum computing depends on the context.
We provide optimal layout synthesis without any ancillary SWAPs as an option.
Note that the resulting encoding may require more SWAP gates than when allowing ancillary qubits.
This option can be encoded in our Two-Way SAT encoding, by simply restricting the SWAP gates
to cases where both the physical qubits are already mapped. 

\subsection{Design Choices}
\label{subsec:designchoices}

\paragraph*{Redundant Cardinality Constraints for Mapping Variables}
Specifying Exactly-One constraints (EO) on logical qubits and At-Most-One constraints (AO) on physical qubits
in the initial time step would be sufficient for correctness.
Note that once the mapping variables are set in the initial time step, the information on bijectivity is propagated
to next time steps, based on the chosen SWAP variables.
However, observe that unrelated to the choice of SWAP variables, some invariants apply to mapping variables in all time steps.
Essentially, the EO and AO constraints are orthogonal to the SWAP variable assignment.
We observed that adding such redundant constraints at each time step
significantly improved solving times. Apparently, this local information can be exploited by the SAT solver during clause learning or unit propagation.
Since we added mostly binary clauses, we conjecture that the improved solving time is due to improved unit propagation.

\paragraph*{Two-Way Encoding vs Explicit CNOT Constraints}
In this paper, we encoded CNOT constraints using a Two-Way encoding instead of explicit CNOT constraints.
We chose Two-Way encoding for two main reasons.
First, we encode transitive closure for predecessors/successors of CNOT gates.
Explicit constraints for CNOT dependencies result in two challenges:
\begin{itemize}
  \item Specifying that the predecessors (successors) of a CNOT gate can not be scheduled in later (earlier)
  time steps results in a quadratic blow-up in the CNOT gates.
  \item Specifying that the predecessors (successors) of a CNOT gate must be scheduled in earlier (later)
  time steps results in long clauses.
\end{itemize}
On the other hand, the Two-Way encoding expresses this bidirectional propagation implicitly
using clauses linear in the number of CNOT gates.
Second, in incremental solving, in each iteration, we need to specify that the goal is reached in the final time step.
Using the Two-Way encoding, we can simply specify that no CNOTs are delayed in the final step.
If we encoded the CNOT constraints explicitly, we would need to specify that every CNOT is scheduled at some time step.
To avoid large clauses, one would need to use auxiliary variables similar to advanced/delayed variables to keep track of the scheduled CNOTs across time steps.
In our tool, we provide the encoding with explicit CNOT constraints as an option.

\section{Experimental Evaluation}
\label{sec:experimentalevaluation}

\subsection{Experiment Design}
\label{subsec:design}
We have extended our tool Q-Synth v1 (Quantum Synthesizer) to include the Two-Way Parallel SAT encoding.
We provide an open-source tool Q-Synth2\footnote{Q-Synth v2 tool with source code, benchmarks, and scripts~\url{https://github.com/irfansha/Q-Synth}} that implements the SAT encoding and the additional options.
For any option chosen, our tool synthesizes a mapped circuit with the (provably) optimal number of additional SWAP+bridge gates.
We use pysat~\cite{imms-sat18} for generating and solving SAT instances incrementally.
For cardinality constraints, we use the sequential counter from pysat.
As a backend for our experiments, we use Cadical-1.53~\cite{BiereFazekasFleuryHeisinger-SAT-Competition-2020-solvers}, a state-of-the-art SAT solver.
One can easily experiment with other SAT solvers in our tool using the pysat interface.
When optimizing the SWAP count, our tool refutes all $k-1$ SWAP+bridge additions if $k$ is optimal.
We report a timeout if an optimal solution is not found within the time limit.
We check equivalence between the original circuits and our mapped circuits with QCEC\footnote{Munich Quantum Toolkit QCEC \url{https://github.com/cda-tum/mqt-qcec}} \cite{Burgholzer_2021_QCEC} for correctness.

We design 3 experiments. Our goal is to investigate the effectiveness of our SAT encoding compared to the current leading tools.
We also compare various additional techniques discussed in~\ref{subsec:additinalfunc}.
For comparison, we consider state-of-the-art tools TB-OLSQ2 (near-optimal)~\cite{OLSQ2_2023} and Qiskit's SABRE (heuristic).
For TB-OLSQ2, we enable the best options i.e., SWAP optimization and upper bound computation by SABRE, with z3 (v4.12.1.0)~\cite{DBLP:conf/tacas/MouraB08} as the backend.
TB-OLSQ2 can provide intermediate non-optimal results. We only report the final (near-optimal) solution when it terminates.
If the tool does not terminate within the time limit, we report it as a timeout.
For SABRE, we use the first 1000 seeds for the SABRE layout and take the minimum SWAPs generated by any seed.
We also compare our results with other leading tools in Section~\ref{sec:introduction}.

\paragraph*{Experiment 1: Standard Benchmarks on Large Platforms}
We consider the standard benchmarks from papers~\cite{OLSQ2_2023,ShaikvdP2023,DBLP:conf/iccad/TanC20} with 23 instances in total.
The benchmark set contains circuits of up to 54 qubits and 270 CNOT gates.
The circuits are mapped to the current NISQ processors, Sycamore with 54 qubits~\cite{arute2019quantum}, Rigetti with 80 qubits\footnote{Rigetti Computing \url{https://www.rigetti.com}} and Eagle with 127 qubits~\cite{chow2021ibm}.
We compare with the tools TB-OLSQ2 and SABRE, with a time limit of 12000 seconds (3hr 20 minutes) for each instance and an 8 GB memory limit.
\paragraph*{Experiment 2: Deep VQE Benchmarks on Large Platforms}
From our experiments and also consistent with the literature~\cite{OLSQ2_2023},
most of the benchmarks from Experiment 1 need at most 10 SWAPs on standard platforms.
To investigate the performance on deep circuits that need many SWAPs,
we use a set of 10 random circuits composed using operators from the Variational Quantum Eigensolver (VQE) algorithm presented in~\cite{majland2023fermionic}.
Our second benchmark set consists of 10 (8 qubit) circuits with up to 79 CNOT gates.
Due to many interactions between the qubits, the number of SWAPs needed to map onto the standard quantum platforms is high.
We use the same time and memory limits as in Experiment 1.
In both Experiments 1 and 2, we denote a timeout with TO.
Here we focus on comparison with TB-OLSQ2 and report both SWAP count and circuit depth.
\paragraph*{Experiment 3: Effectiveness of Additional Functionality}
In this experiment, we compare 4 combinations of SWAPs (S), bridges (B), and relaxed dependencies (R): 1) S 2) S+B 3) S+R 4) S+B+R.
From our two benchmark sets, we consider all the circuits with 14 or fewer qubits and map them onto the standard Melbourne platform of 14 qubits.
We give a time limit of 600 seconds (or 10 minutes) and an 8 GB of memory.
Of the 24 instances generated, we drop qft\_8 which times out in all 4 combinations.
For the rest of the 23 instances, we report SWAPs+bridges for each combination.
Note that every additional SWAP or bridge adds exactly 3 extra CNOTs to the mapped circuit.

\subsection{Results}
\label{subsec:results}

\begin{table}[htbp]
  \caption{Experiment 1: Number of SWAPs required for mapping circuits with QS2: Q-Synth2 (SWAP-optimal), TO2: TB-OLSQ2 (near optimal), and SB: SABRE (heuristic) tools on different platforms.
  Syc: Sycamore (54), Rig: Rigetti (80), Eagle (127) and label or(3/6) represents a circuit "or" with 3 qubits and 6 CNOT gates}
  \label{tb:experiment1}
  \begin{tabular}{lrrrrrrrrr}
      \toprule
      platform (qubits): & \multicolumn{3}{c}{Sycamore (54)} & \multicolumn{3}{c}{Rigetti (80)} & \multicolumn{3}{c}{Eagle (127)}\\
      \cmidrule(lr){2-4}\cmidrule(lr){5-7}\cmidrule(lr){8-10}
      Circuit(q/cx) ~/~ Tool & \textbf{QS2} & TO2 & SB& \textbf{QS2} & TO2 & SB& \textbf{QS2} & TO2 & SB\\
      \hline
      or(3/6)                  & 2           & 2         & 3   &   2          & 2  & 2   & 2           & 2  & 2 \\
      adder(4/10)              & 0           & 0         & 0   &   0          & 0  & 0   & 2           & 2  & 2 \\
      qaoa5(5/8)               & 0           & 0         & 1   &   0          & 0  & 0   & 0           & 0  & 1 \\
      4mod5-v1\_22(5/11)       & 3           & 3         & 4   &   3          & 3  & 5   & 3           & 3 & 6 \\
      mod5mils\_65(5/16)       & 6           & 6         & 7   &   6          & 6  & 7   & 6           & 6  & 8 \\
      4gt13\_92(5/30)          & 10          & 10        & 15  &   10         & 10 & 15  & \textbf{13} & TO & 15 \\
      tof\_4(7/22)             & 1           & 1         & 3   &   1          & 1  & 11  & 3           & 3   & 5 \\
      barenco\_tof4(7/34)      & 5           & 5         & 18  &   6          & 6  & 17  & 8           & 8  & 17 \\
      qft\_8(8/56)             & \textbf{9}  & TO        & 15  &   TO         & TO & 12  & TO          & TO & 23 \\
      tof\_5(9/30)             & 1           & 1         & 3   &   1          & 1  & 5   & 3           & 3   & 12 \\
      mod\_mult55(9/40)        & 6           & 6         & 9   &   \textbf{7} & 8  & 16  & \textbf{12} & TO & 20 \\
      barenco\_tof5(9/50)      & 6           & 6         & 10  &   8          & 8  & 19  & \textbf{12} & TO & 20 \\
      vbe\_adder3(10/50)       & 7           & 7         & 8   &   8          & 8  & 14  & 10          & 10   & 33 \\
      rc\_adder6(14/71)        & TO          & \textbf{8}& 16  &   8          & 8  & 35  & TO          & TO & 51 \\
      ising\_model10(16/90)    & 0           & 0         & 0   &   0          & 0  & 0   & 0           & 0  & 0 \\
      queko(16/15)             & 0           & 0         & 1   &   0          & 0  & 2   & 0           & 0 & 0 \\
      queko(16/29)             & 0           & 0         & 5   &   \textbf{0} & 1  & 12  & 2           & 2 & 14 \\
      queko(16/44)             & 0           & 0         & 7   &   \textbf{0} & 1  & 25  & 2           & 2 &  37\\
      queko(16/58)             & 0           & 0         & 12  &   \textbf{0} & 1  & 20  & \textbf{4}  & TO & 41 \\
      queko(16/87)             & \textbf{0}  & 1         & 10  &   \textbf{0} & 1  & 30  & \textbf{4}  & TO & 36 \\
      queko(16/101)            & 0           & 0         & 18  &   \textbf{0} & 1  & 43  & TO          & TO & 36 \\
      queko(54/54)             & \textbf{0}  & 1         & 12  &   1          & 1  & 31  & TO          & TO & 47 \\
      queko(54/270)            & \textbf{0}  & 1         & 183 &   TO         & TO & 302 & TO          & TO & 428 \\
      \hline
      Total solved of 23       & 22           & 22       & 23   & 21          &  21 &  23 & \textbf{18}& 13 & 23\\
      \bottomrule
  \end{tabular}
  \end{table}
    
\begin{figure}[htbp]
      \centering
      \begin{subfigure}[t]{0.5\textwidth}
        \centering
        \includegraphics[scale=0.43]{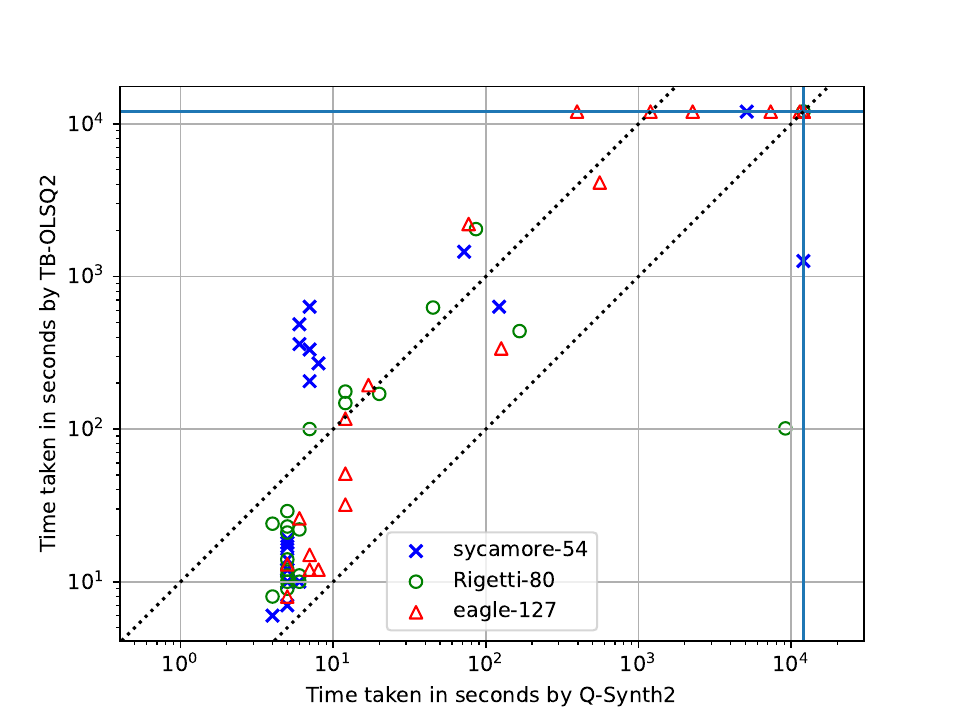}
        \caption{Experiment 1}
        \label{fig:experiment1}
      \end{subfigure}%
      \begin{subfigure}[t]{0.5\textwidth}
        \centering
        \includegraphics[scale=0.43]{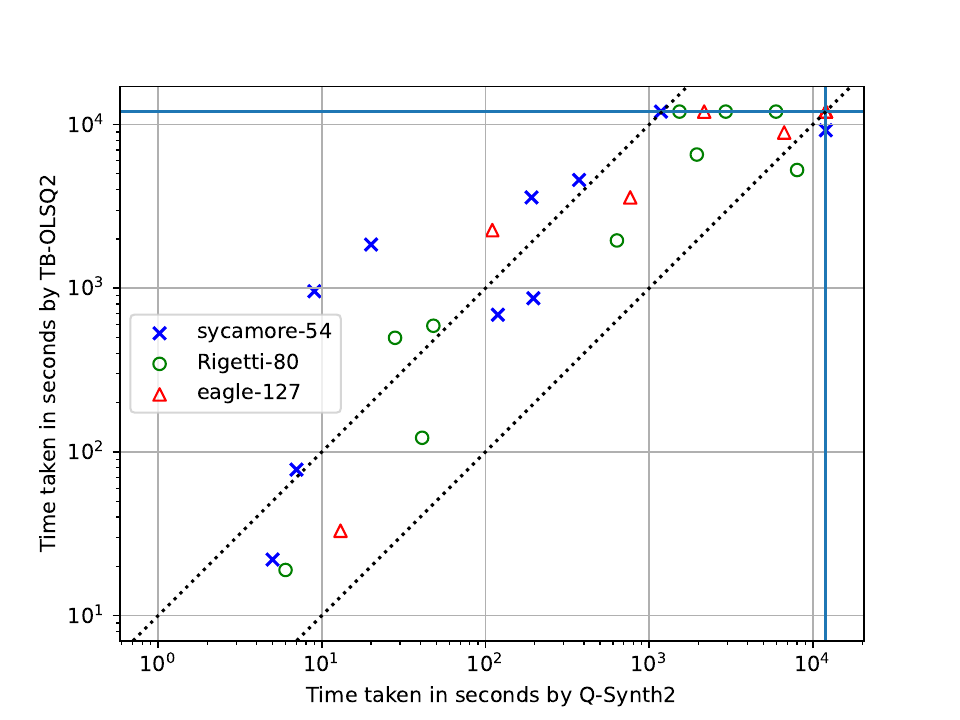}
        \caption{Experiment 2}
        \label{fig:experiment2}
      \end{subfigure}
      \caption{Scatter plots of time taken by TB-OLSQ2 and Q-Synth2.}
      \label{fig:experiment_time}
    \end{figure}

\paragraph*{Experiment 1}
Table \ref{tb:experiment1} reports the number of SWAPs added.
Both on Sycamore and Rigetti, TB-OLSQ2 mostly reports optimal SWAP count (while not proving optimality).
There are 10 instances where it reports near-optimal solutions i.e., only 1 extra SWAP gate or times out.
The difference is more significant on the larger 127-qubit Eagle platform.
Q-Synth2 solved 5 more instances optimally where TB-OLSQ2 times out.
Figure \ref{fig:experiment1} provides the scatter plot of the time taken by Q-Synth2 and TB-OLSQ2.
Except for two instances with rcadder6 (14 qubits) on Sycamore and Rigetti, we significantly outperform TB-OLSQ2 on all three platforms.
In several instances, Q-Synth2 is one or two orders of magnitude faster while proving optimality.
In the case of two instances with rcadder6, being a 14 qubit circuit, our tool takes time to refute the k-1 number of optimal SWAPs.
While the heuristic tool SABRE always returns a mapping within 2 minutes, it also adds too many additional SWAPs.
This observation is consistent with the literature~\cite{OLSQ2_2023}.

\begin{table}[htbp]
  \caption{Experiment 2: Additional SWAPs (s.) and Depth (d.) of mapped VQE circuits on different platforms with QS2: Q-Synth2 (SWAP-optimal) and TO2: TB-OLSQ2 (near optimal)}
  \label{tb:experiment2}
  \centering
  \begin{tabular}{lrrrrrrrrrrrr}
    \toprule
    platform & \multicolumn{4}{c}{Syc (54)} & \multicolumn{4}{c}{Rig (80)} & \multicolumn{4}{c}{Eagle (127)}\\
                           \cmidrule(lr){2-5}\cmidrule(lr){6-9}\cmidrule(lr){10-13}
                  & \multicolumn{2}{c}{\textbf{QS2}} & \multicolumn{2}{c}{TO2} & \multicolumn{2}{c}{\textbf{QS2}} & \multicolumn{2}{c}{TO2} & \multicolumn{2}{c}{\textbf{QS2}} & \multicolumn{2}{c}{TO2}\\
                  \cmidrule(lr){2-3} \cmidrule(lr){4-5}\cmidrule(lr){6-7}\cmidrule(lr){8-9}\cmidrule(lr){10-11}\cmidrule(lr){12-13}
    Circuit(q/cx)              & s.         & d.  & s.         & d.  & s.          & d.  & s.  & d.  & s.          & d. & s. & d.\\
    \midrule
    vqe(8/18)                  & 2           &  34          & 2          & \textbf{33}  & 2           &  36          & 2          &  \textbf{33} & 3           & 38 & 3          & \textbf{35} \\
    vqe(8/39)                  & \textbf{4}  &  65          & 5          & \textbf{62}  & \textbf{6}  &  68          & 7          &  \textbf{63} & \textbf{7}  & 65 & 9          & \textbf{64} \\
    vqe(8/40)                  & \textbf{6}  &  70          & 7          & \textbf{67}  & 7           &  \textbf{67} & 7          &  69          & \textbf{10} & 76 & 12         & \textbf{68} \\
    vqe(8/47)                  & \textbf{8}  &  85          & 10         & \textbf{83}  & 10          &  84         & 10          &  \textbf{83} & \textbf{14} & 92 & 17         & \textbf{86} \\
    vqe(8/48)                  & 6           &  90          & 6          & \textbf{84}  & 8           &  94         & 8           &  \textbf{89} & \textbf{12} & 94 & TO         & TO \\
    vqe(8/52)                  & \textbf{9}  &  90          & 11         & \textbf{87}  & \textbf{11} &  90         & TO          &  TO          & TO          & TO & TO         & TO \\
    vqe(8/63)                  & \textbf{10} & 101          & 11         & \textbf{99}  & 13          & 102         & 13          & 102          & TO          & TO & TO         & TO \\
    vqe(8/71)                  & TO          &  TO          & \textbf{16}& \textbf{111} & \textbf{15} & \textbf{112}& 18          & 113          & TO          & TO & TO         & TO \\
    vqe(8/78)                  & \textbf{14} & \textbf{129} & TO         & TO           & \textbf{17} & \textbf{136}& TO          & TO           & TO          & TO & TO         & TO \\
    vqe(8/79)                  & 11          & 149          & 11         & \textbf{146} & \textbf{16} & \textbf{151}& TO          & TO           & TO          & TO & TO         & TO \\
    \midrule
    Solved (10)                & \multicolumn{2}{c}{9}      & \multicolumn{2}{c}{9}     & \multicolumn{2}{c}{\textbf{10}} &\multicolumn{2}{c}{7} & \multicolumn{2}{c}{\textbf{5}}& \multicolumn{2}{c}{4}  \\
    \bottomrule
  \end{tabular}
  \end{table}

\paragraph*{Experiment 2}
Table \ref{tb:experiment2} reports the number of SWAPs added and circuit depth for the mapped VQE circuits.
On all platforms, Q-Synth2 solves 15 more instances SWAP-optimally out of the 24 instances solved by either of the tools.
TB-OLSQ2 in general reports better circuit depth compared to Q-Synth2.
Interestingly in two instances, vqe(8/40) and vqe(8/71), Q-Synth2 reports better circuit depth.
This shows that TB-OLSQ2 is near optimal in both SWAP additions and circuit depth.
Overall Q-Synth2 also reports near-optimal depth while optimizing additional SWAPs.
Figure \ref{fig:experiment2} shows the scatter plot of time for Experiment 2.
Except for two instances with vqe(8/71) on Sycamore and Rigetti, Q-Synth2 significantly outperforms TB-OLSQ2.

\begin{table}[htbp]
  \caption{Experiment 3: Number of SWAPs+bridges required for mapping deep VQE circuits on Melbourne platform (14-qubits) in 600 seconds with Q-Synth2 with combinations of
           S: Swaps, B: bridges, and R: relaxed dependencies}
  \label{tb:experiment3}
  \begin{tabular}[t]{lrrrr}
    \toprule
      Circuit(q/cx) & {S}  & {SB}  & {SR}    & {SBR}\\
      \midrule
      or(3/6)                 & 2     & 2         &\textbf{1}&   \textbf{1}\\
      adder(4/10)              & 0     & 0         &   0      &   0\\
      qaoa5(5/8)              & 0     & 0         &    0     &   0\\
      4mod5\_22(5/11)       & 3     & \textbf{2}&\textbf{2}&   \textbf{2}\\
      mod5mils65(5/16)       & 6     &\textbf{4} &\textbf{4}&   \textbf{4}\\
      4gt13\_92(5/30)          & 10    &\textbf{8} &\textbf{8}&   \textbf{8}\\
      tof\_4(7/22)             & 1     & 1         &   1      &   1\\
      barencof4(7/34)    & 5     &5          &   5      &   5\\
      tof\_5(9/30)             & 1     &1          &    1     &   1\\
      modmult55(9/40)      & 7     &7          &    7     &   7\\
      barencof5(9/50)   & 6     &6          &    6     &   6\\
      vbe\_adder(10/50)     & 8     &8          &\textbf{6}&   \textbf{6}\\
      rcadder6(14/71)     & 9     &\textbf{8} &    9     &   \textbf{8}\\
      \bottomrule
  \end{tabular}\quad
  \begin{tabular}[t]{lrrrr}
    \toprule
      Circuit(q/cx) & {S}  & {SB}  & {SR}    & {SBR}\\
      \midrule
      vqe(8/18)             & 2     & 2         &   2      &   2\\
      vqe(8/39)             & 6     & 6         &   6      &   6\\
      vqe(8/40)             & 7     & 7         &   7      &   \textbf{6}\\
      vqe(8/47)            & 8     & 8         &   8      &   8\\
      vqe(8/48)             & 7     &\textbf{6} &   7      &   \textbf{6}\\
      vqe(8/52)             & 10    & 10        &   10     &   10\\
      vqe(8/63)            & 12    & 12        &   12     &   12\\
      vqe(8/71)            & 13    &\textbf{12}&   13     &   \textbf{12}\\
      vqe(8/78)            & 17    &15         &   16     &   \textbf{14}\\
      vqe(8/79)            & 15    &\textbf{13}&   14     &   \textbf{13}\\    
      \bottomrule
  \end{tabular}

  \end{table}

\paragraph*{Experiment 3}
Table \ref{tb:experiment3} reports the number of SWAPs+bridges on the Melbourne platform, with a time out of 10 minutes.
Both bridges and relaxed dependencies can reduce the optimal SWAP+bridge additions.
We observe that both techniques together can reduce CNOT count further.
For instance, vqe(8/78) only needs 14 SWAPs+bridges i.e., 9 fewer CNOTs compared to only adding SWAPs.
If we drop any of the options, the optimal CNOT count is higher.

\subsection{Discussion}
\label{sec:discussion}

\paragraph*{Comparison to OLSQ, OLSQ2 and TB-OLSQ2}
In~\cite{OLSQ2_2023}, authors showed that TB-OLSQ2 significantly outperforms both OLSQ and OLSQ2.
Because of the lack of grouping in OLSQ and OLSQ2, the make-span of SMT instances generated is very large.
Since our tool outperforms TB-OLSQ2, we do not report results from the other two directly.
TB-OLSQ2 optimization routines can also be used in our tool to avoid hard unsatisfiable instances when optimality is not needed.

\paragraph*{Comparison to Q-Synth v1 with Classical Planning}
In~\cite{ShaikvdP2023}, we showed that Q-Synth v1 based on Classical Planning outperformed both OLSQ and QMAP.
While the approach scales well for mapping circuits up to 9 qubits onto 14 qubits platforms, larger circuits are out of reach.
For instance, Q-Synth v1 timed out on rcadder6 (Table \ref{tb:experiment3}) with 3 hours.
Q-Synth2 maps the same instance optimally within 5 minutes onto the 14-qubit Melbourne platform.
Q-Synth v1 does not scale well to the other larger quantum platforms.
Mapping individual CNOTs in Q-Synth v1 results in long plan length.
As discussed in the same paper, long plan lengths increase the difficulty of planning.

\paragraph*{Comparison to QMAP and SATMAP}
QMAP~\cite{DBLP:journals/tcad/ZulehnerPW19, DBLP:conf/dac/WilleBZ19} employs an SMT encoding that grows exponentially with the number of physical qubits.
Even using subarchitectures, QMAP is unable to map circuits greater than 7 qubits.
SATMAP~\cite{DBLP:conf/micro/MolaviXDPTA22} on the other hand, encodes the Layout Synthesis as a MAXSAT problem to minimize the number of SWAPs.
It allows the addition of one SWAP before every CNOT and uses MAXSAT solvers to minimize the number of SWAPS.
As shown in~\cite{OLSQ2_2023}, it produces suboptimal solutions and runs out of time even for moderate circuits.

\paragraph*{Comparison with Dynamic Programming Approach}
In~\cite{DBLP:journals/integration/ItokoRIM20}, the authors provided an exact and a heuristic approach for adding SWAPs and bridge gates.
With commutation rules, they showed that using bridges can further reduce the optimal CNOT additions.
Our experiments are consistent with the authors' observations.
Their exact approach already takes 12 minutes to map a 6-qubit circuit to 6-qubit platforms and grows exponentially with the number of qubits.

\paragraph*{Comparison to SABRE}
As observed in  Experiment 1, it is clear that heuristic approaches such as SABRE add too many SWAPs.
Adding many SWAPs not only increases the 2-qubit gate count but also increases circuit depth.
However, heuristic approaches have their place in the quantum compilation pipeline.
As the number of qubits on quantum processors increases, it is necessary to employ a hybrid strategy with heuristic and exact approaches.
For instance, one could use SABRE to quickly find a reasonable initial mapping.
Given such a mapping, one can synthesize a SWAP-optimal mapped circuit using Q-Synth2.
As in TB-OLSQ2, we could use heuristic approaches to get quick upper bounds for near-optimal solving.

\section{Conclusion}

In this paper, we showed that parallel plans can be adapted to preserve SWAP optimality in layout synthesis.
We have encoded the parallel planning problem directly in SAT. We propose a Two-Way encoding, in which
information is propagated both forward and backward, for efficiency.
We also demonstrated that our Two-Way SAT encoding
is compatible with other techniques, like bridges and gate commutation rules.

The technique is implemented in the open-source tool Q-Synth2 for scalable and optimal layout synthesis of deep circuits.
We can optimally map 8-qubit circuits that require up to 14 SWAPs onto a 127-qubit platform.
We significantly outperform leading near-optimal tools while still guaranteeing
that the resulting mapping is optimal.

\newpage
Please cite this paper as:
\begin{verbatim}
@inproceedings{ShaikvdP2024,
  author       = {Irfansha Shaik and Jaco van de Pol},
  title        = {Optimal Layout Synthesis for Deep Quantum Circuits
                  on NISQ Processors with 100+ Qubits},
  booktitle    = {{SAT'24}},
  address      = {Pune, India},
  year         = {2024}
}
\end{verbatim}

\bibliography{references}

\begin{thebibliography}{10}

\bibitem{arute2019quantum}
Frank Arute et~al.
\newblock Quantum supremacy using a programmable superconducting processor.
\newblock {\em Nature}, 574(7779):505--510, 2019.
\newblock \href {https://doi.org/10.1038/s41586-019-1666-5} {\path{doi:10.1038/s41586-019-1666-5}}.

\bibitem{BiereFazekasFleuryHeisinger-SAT-Competition-2020-solvers}
Armin Biere, Katalin Fazekas, Mathias Fleury, and Maximilian Heisinger.
\newblock {CaDiCaL}, {Kissat}, {Paracooba}, {Plingeling} and {Treengeling} entering the {SAT Competition 2020}.
\newblock In {\em Proc.~of {SAT Competition} 2020 -- Solver and Benchmark Descriptions}, volume B-2020-1, pages 51--53. University of Helsinki, 2020.
\newblock URL: \url{https://api.semanticscholar.org/CorpusID:220727106}.

\bibitem{booth2023constraint}
Kyle E.~C. Booth.
\newblock Constraint programming models for depth-optimal qubit assignment and swap-based routing (short paper).
\newblock In {\em 29th International Conference on Principles and Practice of Constraint Programming, {CP} 2023, August 27-31, 2023, Toronto, Canada}, volume 280 of {\em LIPIcs}, pages 43:1--43:10. Schloss Dagstuhl - Leibniz-Zentrum f{\"{u}}r Informatik, 2023.
\newblock \href {https://doi.org/10.4230/LIPICS.CP.2023.43} {\path{doi:10.4230/LIPICS.CP.2023.43}}.

\bibitem{brandhofer2023optimal}
Sebastian Brandhofer, Ilia Polian, and Kevin Krsulich.
\newblock Optimal qubit reuse for near-term quantum computers.
\newblock In {\em {IEEE} International Conference on Quantum Computing and Engineering, {QCE} 2023, Bellevue, WA, USA, September 17-22, 2023}, pages 859--869. {IEEE}, 2023.
\newblock \href {https://doi.org/10.1109/QCE57702.2023.00100} {\path{doi:10.1109/QCE57702.2023.00100}}.

\bibitem{Burgholzer_2021_QCEC}
Lukas Burgholzer and Robert Wille.
\newblock Advanced equivalence checking for quantum circuits.
\newblock {\em IEEE TCAD}, 40(9):1810–1824, 2021.
\newblock \href {https://doi.org/10.1109/tcad.2020.3032630} {\path{doi:10.1109/tcad.2020.3032630}}.

\bibitem{chow2021ibm}
Jerry Chow, Oliver Dial, and Jay Gambetta.
\newblock Ibm quantum breaks the 100-qubit processor barrier.
\newblock {\em IBM Research Blog}, 2, 2021.
\newblock URL: \url{https://www.ibm.com/quantum/blog/127-qubit-quantum-processor-eagle}.

\bibitem{Cross2022}
Andrew Cross, Ali Javadi-Abhari, Thomas Alexander, Lev Bishop, Colm~A. Ryan, Steven Heidel, Niel de~Beaudrap, John Smolin, Jay~M. Gambetta, and Blake~R. Johnson.
\newblock Open quantum assembly language.
\newblock {\em ACM Transactions on Quantum Computing Journal}, 2022.
\newblock URL: \url{https://www.amazon.science/publications/open-quantum-assembly-language}.

\bibitem{DBLP:conf/tacas/MouraB08}
Leonardo~Mendon{\c{c}}a de~Moura and Nikolaj~S. Bj{\o}rner.
\newblock {Z3:} an efficient {SMT} solver.
\newblock In {\em TACAS Proceedings}, LNCS 4963, pages 337--340. Springer, 2008.
\newblock \href {https://doi.org/10.1007/978-3-540-78800-3\_24} {\path{doi:10.1007/978-3-540-78800-3\_24}}.

\bibitem{imms-sat18}
Alexey gnatiev, Antonio Morgado, and Joao Marques-Silva.
\newblock {PySAT:} {A} {Python} toolkit for prototyping with {SAT} oracles.
\newblock In {\em SAT}, pages 428--437, 2018.
\newblock \href {https://doi.org/10.1007/978-3-319-94144-8_26} {\path{doi:10.1007/978-3-319-94144-8_26}}.

\bibitem{DBLP:conf/aspdac/HillmichZW21}
Stefan Hillmich, Alwin Zulehner, and Robert Wille.
\newblock Exploiting quantum teleportation in quantum circuit mapping.
\newblock In {\em {ASPDAC} '21}, pages 792--797. {ACM}, 2021.
\newblock \href {https://doi.org/10.1145/3394885.3431604} {\path{doi:10.1145/3394885.3431604}}.

\bibitem{DBLP:journals/integration/ItokoRIM20}
Toshinari Itoko, Rudy Raymond, Takashi Imamichi, and Atsushi Matsuo.
\newblock Optimization of quantum circuit mapping using gate transformation and commutation.
\newblock {\em Integration}, 70:43--50, 2020.
\newblock \href {https://doi.org/10.1016/j.vlsi.2019.10.004} {\path{doi:10.1016/j.vlsi.2019.10.004}}.

\bibitem{DBLP:conf/aspdac/ItokoRIMC19}
Toshinari Itoko, Rudy Raymond, Takashi Imamichi, Atsushi Matsuo, and Andrew~W. Cross.
\newblock Quantum circuit compilers using gate commutation rules.
\newblock In {\em ASPDAC}, pages 191--196. {ACM}, 2019.
\newblock \href {https://doi.org/10.1145/3287624.3287701} {\path{doi:10.1145/3287624.3287701}}.

\bibitem{DBLP:conf/kr/KautzMS96}
Henry~A. Kautz, David~A. McAllester, and Bart Selman.
\newblock Encoding plans in propositional logic.
\newblock In {\em Proceedings of KR-96}, pages 374--384, November 1996.
\newblock URL: \url{https://henrykautz.com/papers/plankr96.pdf}.

\bibitem{Kautz1992planning}
Henry~A Kautz and Bart Selman.
\newblock Planning as satisfiability.
\newblock In {\em ECAI}, volume~92, pages 359--363, 1992.
\newblock URL: \url{http://www.cs.cornell.edu/selman/papers/pdf/92.ecai.satplan.pdf}.

\bibitem{DBLP:conf/asplos/LiDX19}
Gushu Li, Yufei Ding, and Yuan Xie.
\newblock Tackling the qubit mapping problem for {NISQ}-era quantum devices.
\newblock In {\em {ASPLOS}}, pages 1001--1014. {ACM}, 2019.
\newblock \href {https://doi.org/10.1145/3297858.3304023} {\path{doi:10.1145/3297858.3304023}}.

\bibitem{OLSQ2_2023}
Wan-Hsuan Lin, Jason Kimko, Bochen Tan, Nikolaj Bjørner, and Jason Cong.
\newblock Scalable optimal layout synthesis for {NISQ} quantum processors.
\newblock In {\em DAC}, 2023.
\newblock URL: \url{https://doi.org/10.1109/DAC56929.2023.10247760}.

\bibitem{majland2023fermionic}
Marco Majland, Patrick Ettenhuber, and Nikolaj~Thomas Zinner.
\newblock Fermionic adaptive sampling theory for variational quantum eigensolvers.
\newblock {\em Phys. Rev. A}, 108:052422, Nov 2023.
\newblock \href {https://doi.org/10.1103/PhysRevA.108.052422} {\path{doi:10.1103/PhysRevA.108.052422}}.

\bibitem{CDCL2021}
Joao Marques-Silva, Ines Lynce, and Sharad Malik.
\newblock Conflict-driven clause learning sat solvers.
\newblock {\em Handbook of Satisfiability}, 336:133--182, 2021.
\newblock \href {https://doi.org/10.3233/FAIA200987} {\path{doi:10.3233/FAIA200987}}.

\bibitem{DBLP:conf/micro/MolaviXDPTA22}
Abtin Molavi, Amanda Xu, Martin Diges, Lauren Pick, Swamit Tannu, and Aws Albarghouthi.
\newblock Qubit mapping and routing via {MaxSAT}.
\newblock In {\em MICRO}, pages 1078--1091. {IEEE}, 2022.
\newblock \href {https://doi.org/10.1109/MICRO56248.2022.00077} {\path{doi:10.1109/MICRO56248.2022.00077}}.

\bibitem{Nam_2018}
Yunseong Nam, Neil~J. Ross, Yuan Su, Andrew~M. Childs, and Dmitri Maslov.
\newblock Automated optimization of large quantum circuits with continuous parameters.
\newblock {\em npj Quantum Information}, 4(1), may 2018.
\newblock \href {https://doi.org/10.1038/s41534-018-0072-4} {\path{doi:10.1038/s41534-018-0072-4}}.

\bibitem{Nielsen_Chuang_2010}
Michael~A. Nielsen and Isaac~L. Chuang.
\newblock {\em Quantum circuits}, page 171–215.
\newblock Cambridge University Press, 2010.
\newblock \href {https://doi.org/10.1017/CBO9780511976667.008} {\path{doi:10.1017/CBO9780511976667.008}}.

\bibitem{Peham_2023}
Tom Peham, Lukas Burgholzer, and Robert Wille.
\newblock On optimal subarchitectures for quantum circuit mapping.
\newblock {\em {ACM} Trans. on Quant. Computing}, 2023.
\newblock \href {https://doi.org/10.1145/3593594} {\path{doi:10.1145/3593594}}.

\bibitem{RINTANEN20061031}
Jussi Rintanen, Keijo Heljanko, and Ilkka Niemel{\"{a}}.
\newblock Planning as satisfiability: parallel plans and algorithms for plan search.
\newblock {\em Artif. Intell.}, 170(12-13):1031--1080, 2006.
\newblock \href {https://doi.org/10.1016/J.ARTINT.2006.08.002} {\path{doi:10.1016/J.ARTINT.2006.08.002}}.

\bibitem{ShaikvdP2023}
Irfansha Shaik and Jaco van~de Pol.
\newblock Optimal layout synthesis for quantum circuits as classical planning.
\newblock In {\em {IEEE/ACM} International Conference on Computer Aided Design, {ICCAD} 2023, San Francisco, CA, USA, October 28 - Nov. 2, 2023}, pages 1--9. {IEEE}, 2023.
\newblock \href {https://doi.org/10.1109/ICCAD57390.2023.10323924} {\path{doi:10.1109/ICCAD57390.2023.10323924}}.

\bibitem{DBLP:conf/glvlsi/SrivastavaLCAB23}
Amisha Srivastava, Chao Lu, Navnil Choudhury, Ayush Arunachalam, and Kanad Basu.
\newblock Search space reduction for efficient quantum compilation.
\newblock In {\em Proceedings of {GLSVLSI}-23}, pages 109--114. {ACM}, 2023.
\newblock \href {https://doi.org/10.1145/3583781.3590223} {\path{doi:10.1145/3583781.3590223}}.

\bibitem{DBLP:conf/iccad/TanC20}
Bochen Tan and Jason Cong.
\newblock Optimal layout synthesis for quantum computing.
\newblock In {\em {IEEE/ACM} {ICCAD}}, pages 137:1--137:9. {IEEE}, 2020.
\newblock \href {https://doi.org/10.1145/3400302.3415620} {\path{doi:10.1145/3400302.3415620}}.

\bibitem{DBLP:journals/tc/TanC21}
Bochen Tan and Jason Cong.
\newblock Optimality study of existing quantum computing layout synthesis tools.
\newblock {\em {IEEE} Trans. Computers}, 70(9):1363--1373, 2021.
\newblock \href {https://doi.org/10.1109/TC.2020.3009140} {\path{doi:10.1109/TC.2020.3009140}}.

\bibitem{ijcai2017p620}
Davide Venturelli, Minh Do, Eleanor Rieffel, and Jeremy Frank.
\newblock Temporal planning for compilation of quantum approximate optimization circuits.
\newblock In {\em Proceedings of {IJCAI-17}}, pages 4440--4446, 2017.
\newblock \href {https://doi.org/10.24963/ijcai.2017/620} {\path{doi:10.24963/ijcai.2017/620}}.

\bibitem{DBLP:conf/dac/WilleBZ19}
Robert Wille, Lukas Burgholzer, and Alwin Zulehner.
\newblock Mapping quantum circuits to {IBM} {QX} architectures using the minimal number of {SWAP} and {H} operations.
\newblock In {\em {DAC}-19}, page 142. {ACM}, 2019.
\newblock \href {https://doi.org/10.1145/3316781.3317859} {\path{doi:10.1145/3316781.3317859}}.

\bibitem{DBLP:conf/rc/ZulehnerBW19}
Alwin Zulehner, Hartwig Bauer, and Robert Wille.
\newblock Evaluating the flexibility of a* for mapping quantum circuits.
\newblock In {\em Reversible Computation - 11th International Conference, {RC} 2019, Lausanne, Switzerland, June 24-25, 2019, Proceedings}, volume 11497 of {\em Lecture Notes in Computer Science}, pages 171--190. Springer, 2019.
\newblock \href {https://doi.org/10.1007/978-3-030-21500-2\_11} {\path{doi:10.1007/978-3-030-21500-2\_11}}.

\bibitem{DBLP:journals/tcad/ZulehnerPW19}
Alwin Zulehner, Alexandru Paler, and Robert Wille.
\newblock An efficient methodology for mapping quantum circuits to the {IBM} {QX} architectures.
\newblock {\em {IEEE} TCAD ICS}, 38(7):1226--1236, 2019.
\newblock \href {https://doi.org/10.1109/TCAD.2018.2846658} {\path{doi:10.1109/TCAD.2018.2846658}}.

\end{thebibliography}

\end{document}